\documentclass[prl,twocolumn,showpacs]{revtex4}

\usepackage{graphicx, natbib}
\begin{document}

\title{One- and two-photon spectroscopy of a flux qubit coupled to a microscopic defect}

\author{A. Lupa\c scu}
\altaffiliation[Present address: ] {Institute for Quantum Computing and Department of Physics and Astronomy,
University of Waterloo, 200 University Av. West, Waterloo, ON N2L 3G1, Canada.}
\email{alupascu@iqc.ca}
\author{P. Bertet}
\altaffiliation[Present address: ] {Service de Physique de l'Etat Condens\'{e}, DRECAM,
CEA-Saclay,
91191 Gif-sur-Yvette, France.}
\author{E.F.C. Driessen}
\altaffiliation[Present address: ] {Huygens Laboratory, Leiden University, P.O. Box 9504, 2300 RA Leiden, The Netherlands.}
\author{C.J.P.M. Harmans}
\author{J.E. Mooij}
\affiliation{Kavli Institute of Nanoscience, Delft University of
Technology, PO Box 5046, 2600GA Delft, The Netherlands}

\date{ \today}

\begin{abstract}

%ABSTRACT
We observed the dynamics of a superconducting flux qubit coupled to an extrinsic quantum system (EQS). The presence of the EQS is revealed by an anticrossing in the spectroscopy of the qubit. The excitation of a two-photon transition to the third excited state of the qubit-EQS system allows us to extract detailed information about the energy level structure and the coupling of the EQS. We deduce that the EQS is a two-level system, with a transverse coupling to the qubit. The transition frequency and the coupling of the EQS changed during experiments, which supports the idea that the EQS is a two-level system of microscopic origin.
\end{abstract}
\maketitle

Superconducting qubits are artificial quantum systems that consist of microfabricated circuits including Josephson junctions. Research on these systems is motivated both by the perspective of quantum computing~\cite{_devoret_2004_1,_wendin_2005_1} and by the fact that they are model systems for fundamental studies in quantum mechanics \cite{wallraff_2004_1,chiorescu_2004_1,bertet_2005_1,valenzuela_2006_cooling,lupascu_2007_1}. Decoherence of superconducting qubits is an example of such a topic, relevant both for quantum computing and for understanding the dynamics of open quantum systems.

We report experiments on a superconducting flux qubit, where spectroscopic measurements show that the qubit is coupled to an extrinsic quantum system (EQS). Similar observations have been reported for superconducting phase qubits~\cite{simmonds_2004_1,cooper_2004_1,martinis_2005_1}, where EQSs have been identified as two-level systems (TLS) and showed to cause decoherence of qubits. We study the dynamics of the coupled qubit-EQS system using one-photon spectroscopy, as in~\cite{simmonds_2004_1,cooper_2004_1,martinis_2005_1}, and in addition two-photon spectroscopy. In this letter we show that two-photon spectroscopy provides important information on the energy level structure of the EQS and on its coupling to the qubit. This tool can be used to distinguish a resonance due to a microscopic defect from a spurious resonance in the control or readout circuit, the latter of which can be eliminated by an improved design of these circuits.

The origin of decoherence of superconducting qubits is still not well understood. A few studies have been reported up to date for different types of superconducting qubits~\cite{nakamura_2002_1,duty_2004_2,astafiev_2004_2,bertet_2005_1,yoshihara_2006_1,kakuyanagi_2007_1,ithier_2005_1,metcalfe_2007_DecQuantrBifurcation,simmonds_2004_1,martinis_2005_1}. Decoherence properties are characterized by two different time scales: the energy relaxation time, $T_{1}$, and the dephasing time, $T_{2}$. Systematic studies often show a strong and sample-dependent variation of $T_{1}$ with qubit control parameters~\cite{yoshihara_2006_1}. It is not clear whether relaxation has a microscopic origin or is due to a poorly controlled electromagnetic environment of the qubit. The dephasing time is partly limited by energy relaxation ($T_{2}\leq 2T_{1}$) and is further reduced by slow fluctuations of the qubit parameters, arising from charge, flux, and junction critical-current noise~\cite{ithier_2005_1}. This noise very often has a $1/f$ power spectrum.

Microscopic two-level systems are highly relevant for understanding the decoherence of superconducting qubits. A first reason is that $1/f$ noise is believed to be generated by  a collection of TLSs~\cite{vanharlingen_2004_1}. The second reason is that some TLSs may have an energy level splitting close to the transition frequency of the qubit, thus inducing qubit relaxation. The latter situation was studied in detail for phase qubits. The qubit spectrum displayed a large number of spurious anticrossings~\cite{simmonds_2004_1} that were interpreted as being due to the resonant coupling of the qubit to microscopic TLSs. It was shown~\cite{simmonds_2004_1,martinis_2005_1} that these TLS play a role in decoherence of phase qubits. The coherent exchange of energy between a phase qubit and a TLS was also observed~\cite{cooper_2004_1}.

The fact that many TLSs are observed for phase qubits is generally attributed to the presence of these TLSs inside the barrier of the relatively large-area Josephson junctions (typically 10~$\mu$m$^{2}$~\cite{simmonds_2004_1}) used for phase qubits. Flux, charge-phase, and charge qubits have much smaller Josephson junctions (0.01-0.1~$\mu$m$^{2}$), which could statistically explain the absence of TLSs in many measurements. The presence of coupled TLSs was however observed for small junction qubits (see \emph{e.g.}~\cite{ithier_2005_2} for charge-phase qubits and~\cite{deppe_2007_DynFluxQbCapReadout} for flux qubits). In this paper we report on detailed measurements of a TLS coupled to a flux qubit. We stress that coupled TLS are observed relatively rarely in our experiments, done on one- or two- qubit samples; in addition to the measurements presented here, we observed a TLS in only one other sample. Nevertheless, the frequency of occurrence of such coupled TLSs will become significant in many-qubit samples, even for the case of superconducting qubits with small junctions.

%INSERTING FIG 1
\begin{figure}[!]
\includegraphics[width=3.0in]{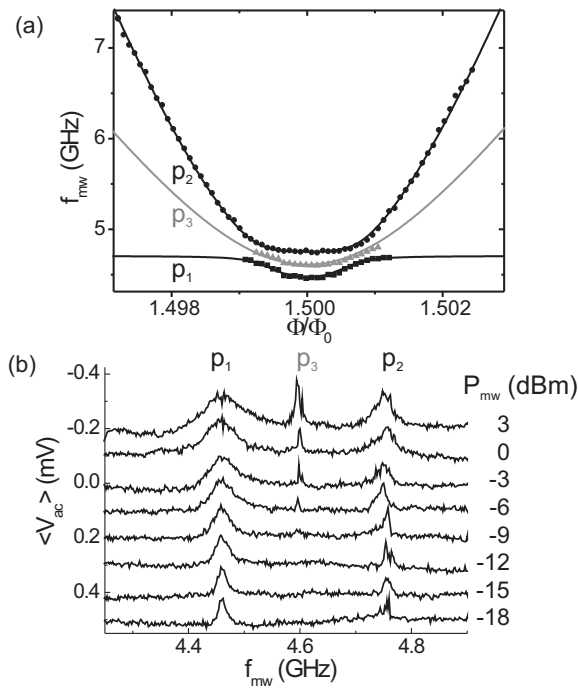}
\caption{\label{fig1} (a) Frequency of the spectroscopy peaks $p_{1}$ (black squares), $p_{2}$ (black circles), and $p_{3}$ (triangles) versus $\Phi$. The black lines are a fit for the peaks $p_{1}$ and $p_{2}$ with the expressions for $E^{qb+TLS}_{01}$ and $E^{qb+TLS}_{02}$, yielding the following parameters: $I_{p}=331$\,nA, $\Delta=4.512$\,GHz, $\nu^{TLS}=4.706$\,GHz, and $g=0.104$\,GHz. The gray line is a plot of $(E^{qb+TLS}_{01}+E^{qb+TLS}_{02})/2$ with the above parameters. (b) Spectroscopy for different values of the microwave power $P_{mw}$ at $\Phi=3\Phi_{0}/2$. The curves are vertically shifted for clarity.}
\end{figure}

The flux qubit~\cite{mooij_1999_1} is formed of a superconducting loop interrupted by three Josephson junctions, two of which are of equal area $S_{J, \mbox{\small{large}}}$ and a third of area $S_{J, \mbox{\small{small}}}=\alpha S_{J, \mbox{\small{large}}}$, where the factor $\alpha\approx $ 0.75. The qubit is fabricated using electron-beam lithography and shadow evaporation of aluminum. The two superconducting aluminum layers that form the junction have thicknesses of $25$\,nm and $50$\,nm. The nominal areas of the qubit junctions are $S_{J, \mbox{\small{large}}}=0.029$\,$\mu$m$^{2}$ and $S_{J, \mbox{\small{small}}}=0.022$\,$\mu$m$^{2}$. The density of the critical current of the junctions is $17$\,$\mu$A$/\mu$m$^{2}$. The qubit control parameter is the magnetic flux applied to the qubit loop, $\Phi$. In the energy eigenbasis, the qubit Hamiltonian is given by
\begin{equation}\label{eq_Hamiltonian_qubit}
 H_{qb}=-{\textstyle\frac{h}{2}} \sqrt{\left[ 2I_{p}/h\left(\Phi-\left ( n+1/2 \right )\Phi_{0}\right ) \right]^{2}+\Delta^{2}}\sigma_{\!z}^{qb} \;,
\end{equation}
where $n$ is the integer part of $\Phi / \Phi_{0}$, $I_{p}$ is the maximum persistent current that flows in the qubit ring, and $\sigma_{\!x,y,z}^{qb}$ are operators that have the Pauli matrices representation in the energy eigenbasis. The parameters $I_{p}$ and $\Delta$ are fixed by design, determined by $S_{J, large}$, $\alpha$, and the critical-current density and capacitance of the Josephson junctions. Transitions between the qubit energy eigenstates are induced by adding to the magnetic flux $\Phi$ a small AC magnetic flux with a frequency resonant with the qubit energy level splitting. The qubit state is measured as follows: a resonant electrical circuit coupled to the qubit is driven with an AC current near resonance, where its impedance is strongly dependent on the state of the qubit. Measuring the voltage $V_{ac}$ across the resonator gives information on the qubit state. This method was described in detail elsewhere~\cite{lupascu_2006_1}.

Spectroscopy is performed by repeating, typically $10^{6}$ times, the following steps~\cite{lupascu_2006_1}: the qubit is first prepared in the ground state by energy relaxation. Transitions to excited states are then induced with microwaves at power $P_{mw}$ and frequency $f_{mw}$, applied for a time $T_{mw}$; for spectroscopy measurements we take $T_{mw}>>T_{1},T_{2}$. As a final step the driving of the resonant circuit used for readout is switched on and the amplitude $V_{ac}$ is measured. The information on the qubit state is provided by the average value of $V_{ac}$, $<V_{ac}>$. In Fig.~\ref{fig1}a the position of the observed spectroscopy peaks for low power is shown as black squares and circles. Away from the symmetry point of the qubit ($\Phi=3\Phi_{0}/2$) the spectrum is similar to the usual flux qubit spectrum: we observe a single peak at frequency $f_{mw}\approx\sqrt{\left[2I_{p}/h\left(\Phi-3\Phi_{0}/2\right)\right]^{2}+\Delta^{2}}$, corresponding to the transition between the ground and excited states of the qubit. However, around $\Phi=3\Phi_{0}/2$, we observe two peaks (labeled $p_{1}$ and $p_{2}$) with a $\Phi$ dependence characteristic of an anticrossing. This reveals the presence of an EQS with a frequency close to the qubit parameter $\Delta$. At larger microwave power a third peak is observed (labeled $p_{3}$) between the peaks $p_{1}$ and $p_{2}$. In Fig.~\ref{fig1}b we plot the average $<V_{ac}>$ as a function of the microwave frequency at $\Phi=3\Phi_{0}/2$, for increasing microwave power.

We start with the reasonable assumption that the EQS is a TLS, following reference~\cite{simmonds_2004_1}. We model the combined qubit-TLS system with the Hamiltonian $H_{qb+TLS}=H_{qb}+H_{TLS}+ H_{qb-TLS}$, with the TLS Hamiltonian
\begin{equation}\label{eq_Hamiltonian_TLS}
  H_{TLS}=-{\textstyle\frac{h}{2}} \nu^{TLS}\sigma_{\!z}^{TLS} \;,
\end{equation}
where $\nu^{TLS}$ is the frequency of the TLS, and the interaction Hamiltonian
\begin{equation}\label{eq_Hamiltonian_interaction}
  H_{qb-TLS}=h g\sigma_{\!x}^{qb} \sigma_{\!x}^{TLS} \;,
\end{equation}
where $\sigma_{\!x,y,z}^{TLS}$ are TLS operators and $g$ is the coupling strength. This Hamiltonian is easily diagonalized, yielding the eigenenergies $E^{qb+TLS}_{n}$ ($n=0$ to $3$) and the transition energies $E^{qb+TLS}_{mn}=E^{qb+TLS}_{n}-E^{qb+TLS}_{m}$. The continuous lines in In Fig.~\ref{fig1}b are a combined fit of the $\Phi$-dependent transition energies $E^{qb+TLS}_{01}$ and $ E^{qb+TLS}_{02}$ with the frequency of the peaks $p_{1}$ and $p_{2}$. This fit yields the parameters $I_{p}$, $\Delta$, $\nu^{TLS}$, and $g$. The agreement of the model with the data is very good. We note that the good agreement does not justify the specific model for the interaction in Eq.~\ref{eq_Hamiltonian_interaction}, as discussed in more detail below, but it justifies the model of resonant interaction and moreover it provides the value of the coupling $g$.

%INSERTING FIG 2
\begin{figure}[!]
\includegraphics[width=3.4 in]{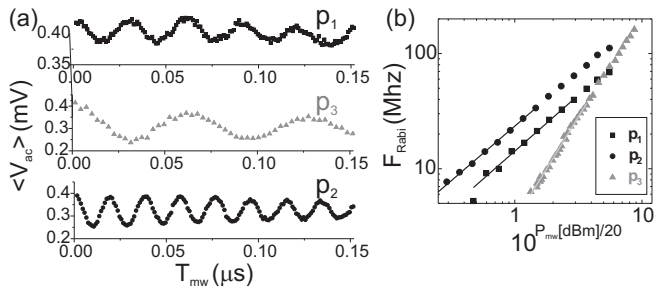}
\caption{\label{fig2}  Measurements of Rabi oscillations at $\Phi=3\Phi_{0}/2$, for a qubit-TLS configuration given by $I_{p}=331$\,nA, $\Delta=4.47$\,GHz, $\nu^{TLS}=4.39$\,GHz, and $g=0.099$\,GHz, for transitions $p_{1}$ (squares), $p_{2}$ (circles), and $p_{3}$ (triangles). (a) Rabi oscillations for microwave power $P_{mw}=7$~dBm. (b) Rabi frequency $F_{\mbox{\small{Rabi}}}$ vs $P_{mw}$. The lines are power law fits for the one- (black) and two-photon (gray) transitions. Only values of $F_{\mbox{\small{Rabi}}}$ smaller than $40$\,MHz are considered for the fit.}
\end{figure}

We now discuss the origin of the peak $p_{3}$ in the spectroscopy signal shown in Fig.~\ref{fig1}. Further understanding on this peak is  provided by the analysis of Rabi oscillations, observed at strong microwave driving. These are shown in Fig.~\ref{fig2}a for three different frequencies, corresponding respectively to the peaks $p_{1}$, $p_{2}$, and $p_{3}$. It is interesting to note that the measurement of the Rabi oscillations shows that the EQS has coherence times comparable to those of the qubit~\cite{zagoskin_2006_1,neeley_2008_process}. The microwave amplitude dependence of the Rabi frequency is shown in Fig.~\ref{fig2}b for the three different transitions. For low microwave power, we observe a power law behavior with exponent $1.0$ for peaks $p_{1}$ and $p_{2}$, and $1.8$ for peak $p_{3}$. This confirms that $p_{1}$ and $p_{2}$ are one-photon transitions. We attribute $p_{3}$ to a two-photon transition to the third excited state of the coupled system. The value of the exponent of the amplitude dependence, $1.8$, is smaller than the ideal value of $2$. This is consistent with numerical simulations of the driven dynamics. We attribute this difference to the partial excitation of the first two excited states of the coupled system.

\begin{figure}[!]
\includegraphics[width=3.4 in]{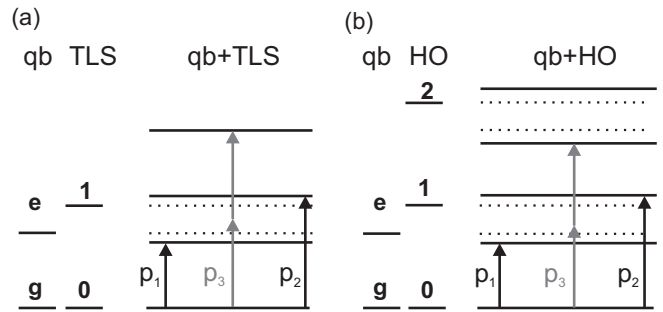}
\caption{\label{fig3} Energy level structure for the qubit (qb) coupled to (a) a two level system (TLS) or (b) an harmonic oscillator (HO). The dotted lines indicates energy levels for the uncoupled system. Black/gray arrows indicate one-/two-photon transitions starting in the ground state and are labeled by $p_{n}$, with $n$ the final state of the coupled system.}
\end{figure}

The observation of the two-photon transition brings important additional information about the coupled EQS. We observe (see Fig.~\ref{fig1}a) that the frequency of the peak $p_{3}$ is the average of the frequencies of peaks $p_{1}$ and $p_{2}$. This is clearly shown by the gray line in Fig.~\ref{fig1}a, which is a plot of the average of the transition energies $E^{qb+TLS}_{01}$ and $E^{qb+TLS}_{02}$, where $E^{qb+TLS}_{01}$ and $E^{qb+TLS}_{02}$ are given by the best fit to the frequencies of $p_{1}$ and $p_{2}$. This particular position of the 2-photon peak is consistent with the hypothesis that the EQS is a TLS , but rules out that the EQS is a HO. This can be understood by considering the structure of levels for the coupled qubit-TLS and qubit-HO cases, as shown in Fig.~\ref{fig3}~a and b respectively. For the latter case the two-photon transition to the third excited state would have a frequency significantly lower than the average value of the one-photon transition frequencies to the first and second excited states. This conclusion holds for any type coupling between the qubit and the HO which is linear in the oscillator creation and annihilation operators. Making the distinction between coupled TLSs and HOs is important since HO modes coupled to the qubit can appear due to spurious resonances in the electromagnetic circuit used to control and read out the qubit.

%INSERTING FIG 3

Having established that the coupled EQS is a TLS, the frequency of the two-photon transition can be used to extract additional details on the type of coupling, as discussed in~\cite{ashhab_2006_RabiTLSQB}. In general, the qubit-TLS coupling can have any terms of the type $h g_{\alpha\beta}\sigma_{\!\alpha}^{qb} \sigma_{\!\beta}^{TLS}$, with $\alpha,\beta=x,y,z$. In our experiment, the two-photon transition frequency is precisely to $(E^{qb+TLS}_{01}+E^{qb+TLS}_{02})/2$, which implies the absence of terms of the type $\sigma_{\!z}^{qb} \sigma_{\!z}^{TLS}$ in the coupling Hamiltonian.

We now discuss the physical origin of the TLS. We first consider the possibility that the TLS is coupled to the qubit by inducing a change in either the critical current or the capacitance of one of the qubit junctions. We calculate numerically the effect of such a change for $\Phi=3\Phi_{0}/2$ and arbitrary values of the offset charges, which are uncontrolled in the experiment (see~\cite{orlando_1999_1} for a model of offset charges). A physically reasonable change of $1\,\%$ in critical current or capacitance result in important changes of the qubit energy-level splitting. However, the corresponding off-diagonal matrix elements in the qubit energy eigenbasis are at least two orders of magnitude too small to explain the value of the coupling $g$ observed in the experiment.

Having ruled out the possibility of a TLS coupled by induced capacitance or critical current changes, we consider now that the qubit couples to the magnetic or the electric field generated by the qubit. Imagine for instance that the TLS is an electronic spin. An estimate for the possible maximum coupling is $g_{m}=\mu_{B}B_{max}$, where $\mu_{B}$ is the Bohr magneton and $B_{max}\approx \mu_{0}I_{p}/\pi t$ is the maximum field generated by the qubit, with $t$ the thickness of the qubit loop lines. For our qubit, with $t=75\,\mbox{nm}$, we estimate $B_{max}\approx 18\,\mbox{mGs}$ resulting in $g_{m}=\mu_{B}B_{max}\approx 25\,\mbox{kHz}$, a value much smaller than observed in the experiment. For electric coupling, we consider now a TLS with an electric-dipole transition of moment $e x_{01}$. The maximum coupling to the qubit is obtained if the TLS is in one of the junction barriers. It has the value $g_{e}=\kappa \Delta\times (x_{01}/d)\times \langle g|\phi_{m}|e\rangle$, where $\kappa=0.5$ or $1$ depending on the TLS being either in one of the two large or in the small qubit junction, $d$ is the barrier thickness, and $\phi_{m}=(\phi_{1}-\phi_{2})/2$ with $\phi_{1}$ and $\phi_{2}$ the phase operators corresponding to the two large qubit junctions. With $\Delta=4.5\,\mbox{GHz}$ and the numerically calculated $\langle g|\phi_{m}|e\rangle=0.77$, we find that $g_{e}$ is equal to the measured value if $\kappa \times (x_{01}/d)=0.03$. Assuming $d=2\,\mbox{nm}$, we obtain $x_{01}=0.06/0.03\,\mbox{nm}$ for the TLS in a large/small qubit junction. This is a physically reasonable value, in agreement with the more systematic studies on phase qubits~\cite{martinis_2005_1}. We thus find that a physically plausible explanation for our experiment is that the TLS is coupled by an electric-dipole transition. Strong coupling through an electric-dipole transition is thus obtained even though our qubit is a flux qubit.

\begin{figure}[!]
\includegraphics[width=3.4 in]{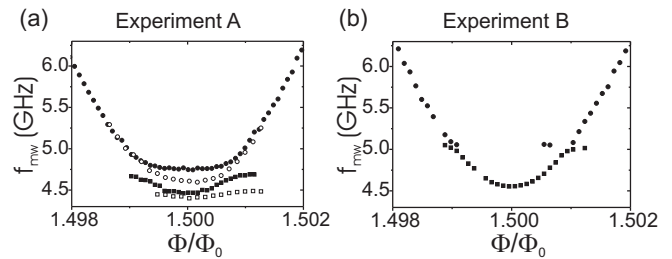}
\caption{\label{fig4} Spectroscopy peaks $p_{1}$ (squares) and $p_{2}$ (circles) for experiment \emph{A} (two different configurations are shown, corresponding to either filled or empty symbols) (a) and for experiment \emph{B} (where a single configuration is observed) (b).}
\end{figure}

During the experiments we observed important changes of the energy-level structure of the combined qubit-TLS system. Our qubit sample was used in two experiments, A and B. Between these two experiments our cryostat was warmed up to room temperature. In experiment A a few different configurations of the energy-level structure were observed. In Fig.~\ref{fig4}a we present the spectroscopy data at small power (only lines $p_{1}$ and $p_{2}$) for two such configurations. The measured spectroscopy is in both cases well described by the qubit-TLS model (given by~Eqs.~\ref{eq_Hamiltonian_qubit}, \ref{eq_Hamiltonian_TLS}, and \ref{eq_Hamiltonian_interaction}), but with different frequency and coupling of the TLS: $\nu^{TLS}=4.706$\,GHz and $g=0.104$\,GHz for the first configuration and $\nu^{TLS}=4.493$\,GHz and $g=0.099$\,GHz for the second configuration. During the first experiment we observed a few changes between such configurations. The change between two configurations was fast on the time scale of a few tens of minutes, which is the time necessary to acquire the data in order to characterize the spectroscopic structure. Each configuration was in turn stable over times of the order of days. We observed for each of these configurations the two-photon transition and Rabi oscillations on all the three transitions. In experiment B we observed a similar spectrum (see Fig.~\ref{fig4}b), with a configuration given by $I_{p}=350$\,nA, $\Delta=4.565$\,GHz, $\nu^{TLS}=5.039$\,GHz, and $g=0.036$\,GHz. The frequency and the coupling of the TLS are significantly different. In contrast to experiment A, the spectrum was stable over all the duration of the experiment (two months). These observations are consistent with other experiments~\cite{simmonds_2004_1,martinis_2008_private} and support the idea that the coupled TLS is of microscopic origin.

%INSERTING FIG 4

In conclusion, we report experiments on a flux qubit coupled to a microscopic two-level system. We stress that the observation of a coupled two-level system is not typical for our measurements on flux qubits. We performed a detailed spectroscopic study of this coupled system. In particular, we used a two-photon transition to the third excited state of the combined system, which is a tool that enabled us to confirm that the coupled system is a two-level system, ruling out the possibility of a coupled harmonic-oscillator mode in the qubit environment. The method described here can be used to shed light on the spectrum of other qubit-EQS systems. We induced coherent transitions to all the three excited states with relatively long coherence times.

%ACKNOWLEDGEMENTS
We thank Franco Nori for comments on the manuscript. This work was supported by the Dutch Organization for Fundamental Research on Matter (FOM), the NanoNed program, and the EU EuroSQIP project.

%REFERENCES
%\bibliography{D:/biblio/physics}

\begin{thebibliography}{27}
\expandafter\ifx\csname natexlab\endcsname\relax\def\natexlab#1{#1}\fi
\expandafter\ifx\csname bibnamefont\endcsname\relax
  \def\bibnamefont#1{#1}\fi
\expandafter\ifx\csname bibfnamefont\endcsname\relax
  \def\bibfnamefont#1{#1}\fi
\expandafter\ifx\csname citenamefont\endcsname\relax
  \def\citenamefont#1{#1}\fi
\expandafter\ifx\csname url\endcsname\relax
  \def\url#1{\texttt{#1}}\fi
\expandafter\ifx\csname urlprefix\endcsname\relax\def\urlprefix{URL }\fi
\providecommand{\bibinfo}[2]{#2}
\providecommand{\eprint}[2][]{\url{#2}}

\bibitem[{\citenamefont{Devoret et~al.}(2004)\citenamefont{Devoret, Wallraff,
  and Martinis}}]{_devoret_2004_1}
\bibinfo{author}{\bibfnamefont{M.}~\bibnamefont{Devoret}},
  \bibinfo{author}{\bibfnamefont{A.}~\bibnamefont{Wallraff}}, \bibnamefont{and}
  \bibinfo{author}{\bibfnamefont{J.}~\bibnamefont{Martinis}},
  \bibinfo{journal}{Arxiv preprint cond-mat/0411174}  (\bibinfo{year}{2004}).

\bibitem[{\citenamefont{Wendin and Shumeiko}(2005)}]{_wendin_2005_1}
\bibinfo{author}{\bibfnamefont{G.}~\bibnamefont{Wendin}} \bibnamefont{and}
  \bibinfo{author}{\bibfnamefont{V.}~\bibnamefont{Shumeiko}},
  \bibinfo{journal}{Arxiv preprint cond-mat/0508729}  (\bibinfo{year}{2005}).

\bibitem[{\citenamefont{Wallraff et~al.}(2004)\citenamefont{Wallraff, Schuster,
  Blais, Frunzio, Huang, Majer, Kumar, Girvin, and
  Schoelkopf}}]{wallraff_2004_1}
\bibinfo{author}{\bibfnamefont{A.}~\bibnamefont{Wallraff}},
  \bibinfo{author}{\bibfnamefont{D.~I.} \bibnamefont{Schuster}},
  \bibinfo{author}{\bibfnamefont{A.}~\bibnamefont{Blais}},
  \bibinfo{author}{\bibfnamefont{L.}~\bibnamefont{Frunzio}},
  \bibinfo{author}{\bibfnamefont{R.~S.} \bibnamefont{Huang}},
  \bibinfo{author}{\bibfnamefont{J.}~\bibnamefont{Majer}},
  \bibinfo{author}{\bibfnamefont{S.}~\bibnamefont{Kumar}},
  \bibinfo{author}{\bibfnamefont{S.~M.} \bibnamefont{Girvin}},
  \bibnamefont{and} \bibinfo{author}{\bibfnamefont{R.~J.}
  \bibnamefont{Schoelkopf}}, \bibinfo{journal}{Nature}
  \textbf{\bibinfo{volume}{431}}, \bibinfo{pages}{162} (\bibinfo{year}{2004}).

\bibitem[{\citenamefont{Chiorescu et~al.}(2004)\citenamefont{Chiorescu, Bertet,
  Semba, Nakamura, Harmans, and Mooij}}]{chiorescu_2004_1}
\bibinfo{author}{\bibfnamefont{I.}~\bibnamefont{Chiorescu}},
  \bibinfo{author}{\bibfnamefont{P.}~\bibnamefont{Bertet}},
  \bibinfo{author}{\bibfnamefont{K.}~\bibnamefont{Semba}},
  \bibinfo{author}{\bibfnamefont{Y.}~\bibnamefont{Nakamura}},
  \bibinfo{author}{\bibfnamefont{C.}~\bibnamefont{Harmans}}, \bibnamefont{and}
  \bibinfo{author}{\bibfnamefont{J.}~\bibnamefont{Mooij}},
  \bibinfo{journal}{Nature} \textbf{\bibinfo{volume}{431}},
  \bibinfo{pages}{159} (\bibinfo{year}{2004}),
  \urlprefix\url{http://dx.doi.org/10.1038/nature02831}.

\bibitem[{\citenamefont{Bertet et~al.}(2005)\citenamefont{Bertet, Chiorescu,
  Burkard, Semba, Harmans, DiVincenzo, and Mooij}}]{bertet_2005_1}
\bibinfo{author}{\bibfnamefont{P.}~\bibnamefont{Bertet}},
  \bibinfo{author}{\bibfnamefont{I.}~\bibnamefont{Chiorescu}},
  \bibinfo{author}{\bibfnamefont{G.}~\bibnamefont{Burkard}},
  \bibinfo{author}{\bibfnamefont{K.}~\bibnamefont{Semba}},
  \bibinfo{author}{\bibfnamefont{C.~J. P.~M.} \bibnamefont{Harmans}},
  \bibinfo{author}{\bibfnamefont{D.~P.} \bibnamefont{DiVincenzo}},
  \bibnamefont{and} \bibinfo{author}{\bibfnamefont{J.~E.} \bibnamefont{Mooij}},
  \bibinfo{journal}{Phys. Rev. Lett.} \textbf{\bibinfo{volume}{95}},
  \bibinfo{pages}{257002} (\bibinfo{year}{2005}).

\bibitem[{\citenamefont{Valenzuela et~al.}(2006)\citenamefont{Valenzuela,
  Oliver, Berns, Berggren, Levitov, and Orlando}}]{valenzuela_2006_cooling}
\bibinfo{author}{\bibfnamefont{S.~O.} \bibnamefont{Valenzuela}},
  \bibinfo{author}{\bibfnamefont{W.~D.} \bibnamefont{Oliver}},
  \bibinfo{author}{\bibfnamefont{D.~M.} \bibnamefont{Berns}},
  \bibinfo{author}{\bibfnamefont{K.~K.} \bibnamefont{Berggren}},
  \bibinfo{author}{\bibfnamefont{L.~S.} \bibnamefont{Levitov}},
  \bibnamefont{and} \bibinfo{author}{\bibfnamefont{T.~P.}
  \bibnamefont{Orlando}}, \bibinfo{journal}{Science}
  \textbf{\bibinfo{volume}{314}}, \bibinfo{pages}{1589} (\bibinfo{year}{2006}).

\bibitem[{\citenamefont{Lupascu et~al.}(2007)\citenamefont{Lupascu, Saito,
  Picot, de~Groot, Harmans, and Mooij}}]{lupascu_2007_1}
\bibinfo{author}{\bibfnamefont{A.}~\bibnamefont{Lupascu}},
  \bibinfo{author}{\bibfnamefont{S.}~\bibnamefont{Saito}},
  \bibinfo{author}{\bibfnamefont{T.}~\bibnamefont{Picot}},
  \bibinfo{author}{\bibfnamefont{P.~C.} \bibnamefont{de~Groot}},
  \bibinfo{author}{\bibfnamefont{C.~J. P.~M.} \bibnamefont{Harmans}},
  \bibnamefont{and} \bibinfo{author}{\bibfnamefont{J.~E.} \bibnamefont{Mooij}},
  \bibinfo{journal}{Nature Physics} \textbf{\bibinfo{volume}{3}},
  \bibinfo{pages}{119} (\bibinfo{year}{2007}).

\bibitem[{\citenamefont{Simmonds et~al.}(2004)\citenamefont{Simmonds, Lang,
  Hite, Nam, Pappas, and Martinis}}]{simmonds_2004_1}
\bibinfo{author}{\bibfnamefont{R.~W.} \bibnamefont{Simmonds}},
  \bibinfo{author}{\bibfnamefont{K.~M.} \bibnamefont{Lang}},
  \bibinfo{author}{\bibfnamefont{D.~A.} \bibnamefont{Hite}},
  \bibinfo{author}{\bibfnamefont{S.}~\bibnamefont{Nam}},
  \bibinfo{author}{\bibfnamefont{D.~P.} \bibnamefont{Pappas}},
  \bibnamefont{and} \bibinfo{author}{\bibfnamefont{J.~M.}
  \bibnamefont{Martinis}}, \bibinfo{journal}{Phys. Rev. Lett.}
  \textbf{\bibinfo{volume}{93}}, \bibinfo{pages}{077003}
  (\bibinfo{year}{2004}).

\bibitem[{\citenamefont{Cooper et~al.}(2004)\citenamefont{Cooper, Steffen,
  McDermott, Simmonds, Oh, Hite, Pappas, and Martinis}}]{cooper_2004_1}
\bibinfo{author}{\bibfnamefont{K.~B.} \bibnamefont{Cooper}},
  \bibinfo{author}{\bibfnamefont{M.}~\bibnamefont{Steffen}},
  \bibinfo{author}{\bibfnamefont{R.}~\bibnamefont{McDermott}},
  \bibinfo{author}{\bibfnamefont{R.~W.} \bibnamefont{Simmonds}},
  \bibinfo{author}{\bibfnamefont{S.}~\bibnamefont{Oh}},
  \bibinfo{author}{\bibfnamefont{D.~A.} \bibnamefont{Hite}},
  \bibinfo{author}{\bibfnamefont{D.~P.} \bibnamefont{Pappas}},
  \bibnamefont{and} \bibinfo{author}{\bibfnamefont{J.~M.}
  \bibnamefont{Martinis}}, \bibinfo{journal}{Phys. Rev. Lett.}
  \textbf{\bibinfo{volume}{93}}, \bibinfo{pages}{180401}
  (\bibinfo{year}{2004}).

\bibitem[{\citenamefont{Martinis et~al.}(2005)\citenamefont{Martinis, Cooper,
  McDermott, Steffen, Ansmann, Osborn, Cicak, Oh, Pappas, Simmonds
  et~al.}}]{martinis_2005_1}
\bibinfo{author}{\bibfnamefont{J.}~\bibnamefont{Martinis}},
  \bibinfo{author}{\bibfnamefont{K.}~\bibnamefont{Cooper}},
  \bibinfo{author}{\bibfnamefont{R.}~\bibnamefont{McDermott}},
  \bibinfo{author}{\bibfnamefont{M.}~\bibnamefont{Steffen}},
  \bibinfo{author}{\bibfnamefont{M.}~\bibnamefont{Ansmann}},
  \bibinfo{author}{\bibfnamefont{K.~D.} \bibnamefont{Osborn}},
  \bibinfo{author}{\bibfnamefont{K.}~\bibnamefont{Cicak}},
  \bibinfo{author}{\bibfnamefont{S.}~\bibnamefont{Oh}},
  \bibinfo{author}{\bibfnamefont{D.}~\bibnamefont{Pappas}},
  \bibinfo{author}{\bibfnamefont{R.}~\bibnamefont{Simmonds}},
  \bibnamefont{et~al.}, \bibinfo{journal}{Phys. Rev. Lett.}
  \textbf{\bibinfo{volume}{93}}, \bibinfo{pages}{210503}
  (\bibinfo{year}{2005}).

\bibitem[{\citenamefont{Nakamura et~al.}(2002)\citenamefont{Nakamura, Pashkin,
  Yamamoto, and Tsai}}]{nakamura_2002_1}
\bibinfo{author}{\bibfnamefont{Y.}~\bibnamefont{Nakamura}},
  \bibinfo{author}{\bibfnamefont{Y.}~\bibnamefont{Pashkin}},
  \bibinfo{author}{\bibfnamefont{T.}~\bibnamefont{Yamamoto}}, \bibnamefont{and}
  \bibinfo{author}{\bibfnamefont{J.}~\bibnamefont{Tsai}},
  \bibinfo{journal}{Phys. Rev. Lett.} \textbf{\bibinfo{volume}{88}},
  \bibinfo{pages}{47901} (\bibinfo{year}{2002}).

\bibitem[{\citenamefont{Duty et~al.}(2004)\citenamefont{Duty, Bladh,
  Gunnarsson, and Delsing}}]{duty_2004_2}
\bibinfo{author}{\bibfnamefont{T.}~\bibnamefont{Duty}},
  \bibinfo{author}{\bibfnamefont{K.}~\bibnamefont{Bladh}},
  \bibinfo{author}{\bibfnamefont{D.}~\bibnamefont{Gunnarsson}},
  \bibnamefont{and} \bibinfo{author}{\bibfnamefont{P.}~\bibnamefont{Delsing}},
  \bibinfo{journal}{Journal of Low Temperature Physics}
  \textbf{\bibinfo{volume}{136}}, \bibinfo{pages}{453} (\bibinfo{year}{2004}).

\bibitem[{\citenamefont{Astafiev et~al.}(2004)\citenamefont{Astafiev, Pashkin,
  Nakamura, Yamamoto, and Tsai}}]{astafiev_2004_2}
\bibinfo{author}{\bibfnamefont{O.}~\bibnamefont{Astafiev}},
  \bibinfo{author}{\bibfnamefont{Y.}~\bibnamefont{Pashkin}},
  \bibinfo{author}{\bibfnamefont{Y.}~\bibnamefont{Nakamura}},
  \bibinfo{author}{\bibfnamefont{T.}~\bibnamefont{Yamamoto}}, \bibnamefont{and}
  \bibinfo{author}{\bibfnamefont{J.}~\bibnamefont{Tsai}},
  \bibinfo{journal}{Physical Review Letters} \textbf{\bibinfo{volume}{93}},
  \bibinfo{pages}{267007} (\bibinfo{year}{2004}).

\bibitem[{\citenamefont{Yoshihara et~al.}(2006)\citenamefont{Yoshihara,
  Harrabi, Niskanen, Nakamura, and Tsai}}]{yoshihara_2006_1}
\bibinfo{author}{\bibfnamefont{F.}~\bibnamefont{Yoshihara}},
  \bibinfo{author}{\bibfnamefont{K.}~\bibnamefont{Harrabi}},
  \bibinfo{author}{\bibfnamefont{A.~O.} \bibnamefont{Niskanen}},
  \bibinfo{author}{\bibfnamefont{Y.}~\bibnamefont{Nakamura}}, \bibnamefont{and}
  \bibinfo{author}{\bibfnamefont{J.~S.} \bibnamefont{Tsai}},
  \bibinfo{journal}{Phys. Rev. Lett.} \textbf{\bibinfo{volume}{97}},
  \bibinfo{pages}{167001} (\bibinfo{year}{2006}).

\bibitem[{\citenamefont{Kakuyanagi et~al.}(2007)\citenamefont{Kakuyanagi, Meno,
  Saito, Nakano, Semba, Takayanagi, Deppe, and Shnirman}}]{kakuyanagi_2007_1}
\bibinfo{author}{\bibfnamefont{K.}~\bibnamefont{Kakuyanagi}},
  \bibinfo{author}{\bibfnamefont{T.}~\bibnamefont{Meno}},
  \bibinfo{author}{\bibfnamefont{S.}~\bibnamefont{Saito}},
  \bibinfo{author}{\bibfnamefont{H.}~\bibnamefont{Nakano}},
  \bibinfo{author}{\bibfnamefont{K.}~\bibnamefont{Semba}},
  \bibinfo{author}{\bibfnamefont{H.}~\bibnamefont{Takayanagi}},
  \bibinfo{author}{\bibfnamefont{F.}~\bibnamefont{Deppe}}, \bibnamefont{and}
  \bibinfo{author}{\bibfnamefont{A.}~\bibnamefont{Shnirman}},
  \bibinfo{journal}{Phys. Rev. Lett.} \textbf{\bibinfo{volume}{98}},
  \bibinfo{pages}{047004} (\bibinfo{year}{2007}).

\bibitem[{\citenamefont{Ithier et~al.}(2005)\citenamefont{Ithier, Collin,
  Joyez, Meeson, Vion, Esteve, Chiarello, Shnirman, Makhlin, Schriefl
  et~al.}}]{ithier_2005_1}
\bibinfo{author}{\bibfnamefont{G.}~\bibnamefont{Ithier}},
  \bibinfo{author}{\bibfnamefont{E.}~\bibnamefont{Collin}},
  \bibinfo{author}{\bibfnamefont{P.}~\bibnamefont{Joyez}},
  \bibinfo{author}{\bibfnamefont{P.}~\bibnamefont{Meeson}},
  \bibinfo{author}{\bibfnamefont{D.}~\bibnamefont{Vion}},
  \bibinfo{author}{\bibfnamefont{D.}~\bibnamefont{Esteve}},
  \bibinfo{author}{\bibfnamefont{F.}~\bibnamefont{Chiarello}},
  \bibinfo{author}{\bibfnamefont{A.}~\bibnamefont{Shnirman}},
  \bibinfo{author}{\bibfnamefont{Y.}~\bibnamefont{Makhlin}},
  \bibinfo{author}{\bibfnamefont{J.}~\bibnamefont{Schriefl}},
  \bibnamefont{et~al.}, \bibinfo{journal}{Phys. Rev. B}
  \textbf{\bibinfo{volume}{72}}, \bibinfo{pages}{134519}
  (\bibinfo{year}{2005}).

\bibitem[{\citenamefont{Metcalfe et~al.}(2007)\citenamefont{Metcalfe, Boaknin,
  Manucharyan, Vijay, Siddiqi, Rigetti, Frunzio, Schoelkopf, and
  Devoret}}]{metcalfe_2007_DecQuantrBifurcation}
\bibinfo{author}{\bibfnamefont{M.}~\bibnamefont{Metcalfe}},
  \bibinfo{author}{\bibfnamefont{E.}~\bibnamefont{Boaknin}},
  \bibinfo{author}{\bibfnamefont{V.}~\bibnamefont{Manucharyan}},
  \bibinfo{author}{\bibfnamefont{R.}~\bibnamefont{Vijay}},
  \bibinfo{author}{\bibfnamefont{I.}~\bibnamefont{Siddiqi}},
  \bibinfo{author}{\bibfnamefont{C.}~\bibnamefont{Rigetti}},
  \bibinfo{author}{\bibfnamefont{L.}~\bibnamefont{Frunzio}},
  \bibinfo{author}{\bibfnamefont{R.~J.} \bibnamefont{Schoelkopf}},
  \bibnamefont{and} \bibinfo{author}{\bibfnamefont{M.~H.}
  \bibnamefont{Devoret}}, \bibinfo{journal}{Physical Review B (Condensed Matter
  and Materials Physics)} \textbf{\bibinfo{volume}{76}}, \bibinfo{eid}{174516}
  (pages~\bibinfo{numpages}{5}) (\bibinfo{year}{2007}),
  \urlprefix\url{http://link.aps.org/abstract/PRB/v76/e174516}.

\bibitem[{\citenamefont{Van~Harlingen et~al.}(2004)\citenamefont{Van~Harlingen,
  Robertson, Plourde, Reichardt, Crane, and Clarke}}]{vanharlingen_2004_1}
\bibinfo{author}{\bibfnamefont{D.}~\bibnamefont{Van~Harlingen}},
  \bibinfo{author}{\bibfnamefont{T.}~\bibnamefont{Robertson}},
  \bibinfo{author}{\bibfnamefont{B.}~\bibnamefont{Plourde}},
  \bibinfo{author}{\bibfnamefont{P.}~\bibnamefont{Reichardt}},
  \bibinfo{author}{\bibfnamefont{T.}~\bibnamefont{Crane}}, \bibnamefont{and}
  \bibinfo{author}{\bibfnamefont{J.}~\bibnamefont{Clarke}},
  \bibinfo{journal}{Physical Review B} \textbf{\bibinfo{volume}{70}},
  \bibinfo{pages}{64517} (\bibinfo{year}{2004}).

\bibitem[{\citenamefont{Ithier}(2005)}]{ithier_2005_2}
\bibinfo{author}{\bibfnamefont{G.}~\bibnamefont{Ithier}}, Ph.D. thesis,
  \bibinfo{school}{CEA Saclay} (\bibinfo{year}{2005}).

\bibitem[{\citenamefont{Deppe et~al.}(2007)\citenamefont{Deppe, Mariantoni,
  Menzel, Saito, Kakuyanagi, Tanaka, Meno, Semba, Takayanagi, and
  Gross}}]{deppe_2007_DynFluxQbCapReadout}
\bibinfo{author}{\bibfnamefont{F.}~\bibnamefont{Deppe}},
  \bibinfo{author}{\bibfnamefont{M.}~\bibnamefont{Mariantoni}},
  \bibinfo{author}{\bibfnamefont{E.~P.} \bibnamefont{Menzel}},
  \bibinfo{author}{\bibfnamefont{S.}~\bibnamefont{Saito}},
  \bibinfo{author}{\bibfnamefont{K.}~\bibnamefont{Kakuyanagi}},
  \bibinfo{author}{\bibfnamefont{H.}~\bibnamefont{Tanaka}},
  \bibinfo{author}{\bibfnamefont{T.}~\bibnamefont{Meno}},
  \bibinfo{author}{\bibfnamefont{K.}~\bibnamefont{Semba}},
  \bibinfo{author}{\bibfnamefont{H.}~\bibnamefont{Takayanagi}},
  \bibnamefont{and} \bibinfo{author}{\bibfnamefont{R.}~\bibnamefont{Gross}},
  \bibinfo{journal}{Physical Review B} \textbf{\bibinfo{volume}{76}},
  \bibinfo{eid}{214503} (pages~\bibinfo{numpages}{19}) (\bibinfo{year}{2007}),
  \urlprefix\url{http://link.aps.org/abstract/PRB/v76/e214503}.

\bibitem[{\citenamefont{Mooij et~al.}(1999)\citenamefont{Mooij, Orlando,
  Levitov, Tian, van~der Wal, , and Lloyd}}]{mooij_1999_1}
\bibinfo{author}{\bibfnamefont{J.~E.} \bibnamefont{Mooij}},
  \bibinfo{author}{\bibfnamefont{T.~P.} \bibnamefont{Orlando}},
  \bibinfo{author}{\bibfnamefont{L.}~\bibnamefont{Levitov}},
  \bibinfo{author}{\bibfnamefont{L.}~\bibnamefont{Tian}},
  \bibinfo{author}{\bibfnamefont{C.~H.} \bibnamefont{van~der Wal}}, ,
  \bibnamefont{and} \bibinfo{author}{\bibfnamefont{S.}~\bibnamefont{Lloyd}},
  \bibinfo{journal}{Science} \textbf{\bibinfo{volume}{285}},
  \bibinfo{pages}{1036} (\bibinfo{year}{1999}).

\bibitem[{\citenamefont{Lupascu et~al.}(2006)\citenamefont{Lupascu, Driessen,
  Roschier, Harmans, and Mooij}}]{lupascu_2006_1}
\bibinfo{author}{\bibfnamefont{A.}~\bibnamefont{Lupascu}},
  \bibinfo{author}{\bibfnamefont{E.~F.~C.} \bibnamefont{Driessen}},
  \bibinfo{author}{\bibfnamefont{L.}~\bibnamefont{Roschier}},
  \bibinfo{author}{\bibfnamefont{C.~J. P.~M.} \bibnamefont{Harmans}},
  \bibnamefont{and} \bibinfo{author}{\bibfnamefont{J.~E.} \bibnamefont{Mooij}},
  \bibinfo{journal}{Phys. Rev. Lett.} \textbf{\bibinfo{volume}{96}},
  \bibinfo{pages}{127003} (\bibinfo{year}{2006}).

\bibitem[{\citenamefont{Zagoskin et~al.}(2006)\citenamefont{Zagoskin, Ashhab,
  Johansson, and Nori}}]{zagoskin_2006_1}
\bibinfo{author}{\bibfnamefont{A.}~\bibnamefont{Zagoskin}},
  \bibinfo{author}{\bibfnamefont{S.}~\bibnamefont{Ashhab}},
  \bibinfo{author}{\bibfnamefont{J.}~\bibnamefont{Johansson}},
  \bibnamefont{and} \bibinfo{author}{\bibfnamefont{F.}~\bibnamefont{Nori}},
  \bibinfo{journal}{Physical review letters} \textbf{\bibinfo{volume}{97}},
  \bibinfo{pages}{77001} (\bibinfo{year}{2006}).

\bibitem[{\citenamefont{Neeley et~al.}(2008)\citenamefont{Neeley, Ansmann,
  Bialczak, Hofheinz, Katz, Lucero, O'Connell, Wang, Cleland, and
  Martinis}}]{neeley_2008_process}
\bibinfo{author}{\bibfnamefont{M.}~\bibnamefont{Neeley}},
  \bibinfo{author}{\bibfnamefont{M.}~\bibnamefont{Ansmann}},
  \bibinfo{author}{\bibfnamefont{R.}~\bibnamefont{Bialczak}},
  \bibinfo{author}{\bibfnamefont{M.}~\bibnamefont{Hofheinz}},
  \bibinfo{author}{\bibfnamefont{N.}~\bibnamefont{Katz}},
  \bibinfo{author}{\bibfnamefont{E.}~\bibnamefont{Lucero}},
  \bibinfo{author}{\bibfnamefont{A.}~\bibnamefont{O'Connell}},
  \bibinfo{author}{\bibfnamefont{H.}~\bibnamefont{Wang}},
  \bibinfo{author}{\bibfnamefont{A.}~\bibnamefont{Cleland}}, \bibnamefont{and}
  \bibinfo{author}{\bibfnamefont{J.}~\bibnamefont{Martinis}},
  \bibinfo{journal}{Nature Physics} \textbf{\bibinfo{volume}{4}},
  \bibinfo{pages}{523} (\bibinfo{year}{2008}).

\bibitem[{\citenamefont{Ashhab et~al.}(2006)\citenamefont{Ashhab, Johansson,
  and Nori}}]{ashhab_2006_RabiTLSQB}
\bibinfo{author}{\bibfnamefont{S.}~\bibnamefont{Ashhab}},
  \bibinfo{author}{\bibfnamefont{J.}~\bibnamefont{Johansson}},
  \bibnamefont{and} \bibinfo{author}{\bibfnamefont{F.}~\bibnamefont{Nori}},
  \bibinfo{journal}{New Journal of Physics} \textbf{\bibinfo{volume}{8}},
  \bibinfo{pages}{103} (\bibinfo{year}{2006}).

\bibitem[{\citenamefont{Orlando et~al.}(1999)\citenamefont{Orlando, Mooij,
  Tian, van~der Wal, Levitov, Lloyd, and Mazo}}]{orlando_1999_1}
\bibinfo{author}{\bibfnamefont{T.~P.} \bibnamefont{Orlando}},
  \bibinfo{author}{\bibfnamefont{J.~E.} \bibnamefont{Mooij}},
  \bibinfo{author}{\bibfnamefont{L.}~\bibnamefont{Tian}},
  \bibinfo{author}{\bibfnamefont{C.~H.} \bibnamefont{van~der Wal}},
  \bibinfo{author}{\bibfnamefont{L.~S.} \bibnamefont{Levitov}},
  \bibinfo{author}{\bibfnamefont{S.}~\bibnamefont{Lloyd}}, \bibnamefont{and}
  \bibinfo{author}{\bibfnamefont{J.~J.} \bibnamefont{Mazo}},
  \bibinfo{journal}{Phys. Rev. B} \textbf{\bibinfo{volume}{60}},
  \bibinfo{pages}{15398} (\bibinfo{year}{1999}).

\bibitem[{\citenamefont{Martinis}()}]{martinis_2008_private}
\bibinfo{author}{\bibfnamefont{J.}~\bibnamefont{Martinis}},
  \bibinfo{note}{private communication}.

\end{thebibliography}

\end{document}